\documentclass[useAMS,usenatbib]{mn2e}
\usepackage{amsmath}
\usepackage{txfonts,amssymb}
\usepackage{graphicx}
\usepackage{hyperref}
\usepackage{afterpage,array,longtable}

\def\asec{\ifmmode ^{\prime\prime}\else$^{\prime\prime}$\fi}
\def\etal{{et\,al. }}
\def\msun{M$_{\odot}$}

\def\degs{\ifmmode ^{\circ}\else$^{\circ}$\fi}
\def\amin{\ifmmode ^{\prime}\else$^{\prime}$\fi}
\def\asec{\ifmmode ^{\prime\prime}\else$^{\prime\prime}$\fi}
\def\farcs{\hbox{$.\!\!^{\prime\prime}$}}   

\def\degs{\ifmmode ^{\circ}\else$^{\circ}$\fi}
\def\amin{\ifmmode ^{\prime}\else$^{\prime}$\fi}

\def\aap{{A\&A\/}}
\def\memsai{Mem.~Soc.~Astron.~Italiana}

\def\eden{\ensuremath{N_{\text{e}}}}

\def\cm3{\rm ~cm^{-3}}
\def\kms{\ensuremath{\,\text{km}\,\text{s}^{-1}}}

\def\Msun{~{\rm M}_\odot}
\def\Ti44{M(^{44}{\rm Ti})}

\def\psr{PSR\,B0540--69.3}
\def\snr{SNR~0540--69.3}
\def\pthreed{\textsc{p3d}}

\def\paperone{L11}
\def\paperser{S05}
\def\pmorse{M06}

\DeclareRobustCommand{\ion}[2]{%
\relax\ifmmode
\ifx\testbx\f@series
{\mathbf{#1\,\mathsc{#2}}}\else
{\mathrm{#1\,\mathsc{#2}}}\fi
\else\textup{#1\,{\mdseries\textsc{#2}}}%
\fi}

\def\lsim{\!\!\!\phantom{\le}\smash{\buildrel{}\over
  {\lower2.5dd\hbox{$\buildrel{\lower2dd\hbox{$\displaystyle<$}}\over
                               \sim$}}}\,\,}
\def\gsim{\!\!\!\phantom{\ge}\smash{\buildrel{}\over
  {\lower2.5dd\hbox{$\buildrel{\lower2dd\hbox{$\displaystyle>$}}\over
                               \sim$}}}\,\,}

\title[The three-dimensional structure of SNR 0540--69.3]
{Properties of the three-dimensional structure in the central region of\\ the supernova remnant SNR 0540--69.3
\thanks{Based on observations collected at the European Organisation for Astronomical Research in the Southern Hemisphere, Chile; proposal number 086.D-0992, P.I.\ P.\ Lundqvist.}}

\author[C.~Sandin et al.]{C.~Sandin$^{1}$\thanks{E-mail: CSandin@aip.de}, P.~Lundqvist$^{2}$, N.~Lundqvist$^{2}$, C.-I.~Bj\"ornsson$^{2}$, G.~Olofsson$^{2}$, 
\newauthor
Yu.~A.~Shibanov$^{3,4}$\\
$^{1}$Leibniz-Institut f\"ur Astrophysik Potsdam (AIP), An der Sternwarte 16, 144 82 Potsdam, Germany\\
$^{2}$Department of Astronomy, Stockholm University, AlbaNova Science Center, Oskar Klein Centre, SE-106 91 Stockholm, Sweden\\
$^{3}$Ioffe Physical Technical Institute, Politekhnicheskaya 26, St. Petersburg, 194021, Russia\\
$^{4}$St. Petersburg State Polytechnical Univ., Polytekhnicheskaya 29, St.\ Petersburg, 195251, Russia\\}

\date{Submitted 12 February 2013}

\pagerange{\pageref{firstpage}--\pageref{lastpage}} \pubyear{2013}

\begin{document}
\label{firstpage}
\maketitle

\begin{abstract}
We present and discuss new visual wavelength-range observations of the inner regions of the supernova remnant {\snr} that is located in the Large Magellanic Cloud (LMC). These observations provide us with more spatial and spectral information than were previously available for this object. We use these data to create a detailed three-dimensional model of the remnant, assuming linear expansion of the ejecta. With the observations and the model we study the general three-dimensional structure of the remnant, and the influence of an active region in the remnant -- a ``blob'' -- that we address in previous papers. We used the fibre-fed integral-field spectrograph VIMOS at the Very Large Telescope of the European Southern Observatory. The observations provide us with three-dimensional data in [\ion{O}{iii}]$\,\lambda5007$ and [\ion{S}{ii}]$\,\lambda\lambda6717,6731$ at an $0\farcs33\!\times\!0\farcs33$ spatial sampling and a velocity resolution of about 35\kms. We decomposed the two, partially overlapping, sulphur lines and used them to calculate electron densities across the remnant at high signal-to-noise ratio. In our study we recover results of previous studies, but we are more importantly able to obtain more detailed information than before. Our analysis reveals a structure that stretches from the position of the ``blob'', and into the plane of the sky at a position angle of PA\,$\simeq\!60${\degs}. Assuming a remnant age of 1000 years and the usual LMC distance, the structure has an inclination angle of about 65{\degs} to the line of sight. The position angle is close to the symmetry axis with present and past activity in the visual and the X-ray wavelength ranges. We speculate that the pulsar is positioned along this activity axis, where it has a velocity along the line of sight of a few hundred \kms. The ``blob'' is most likely a region of shock activity, as it is mainly bright in [\ion{S}{ii}]; future observations of [\ion{O}{ii}]$\,\lambda\lambda3726,3729$ would be useful to test whether the S/O abundance ratio is higher than average for that location in the remnant. The striking resemblance in X-rays between the pulsar wind nebula (PWN) of {\snr} and the Crab, in combination with our findings in this paper, suggests that the symmetry axis is part of a torus in the PWN. This is in agreement with the original suggestion by Gotthelf \& Wang.
\end{abstract}


\section{Introduction}
The 50.2 millisecond pulsar {\psr}, including the surrounding supernova remnant {\snr}, in the Large Magellanic Cloud (LMC), is often referred to as the ``Crab twin''. Its assumed age of about 1000 years, its pulse period, and its surrounding pulsar wind nebula (PWN), are all similar to the Crab. The most obvious differences between the two pulsars and their surrounding nebulae are that the supernova ejecta of {\snr} are oxygen-rich, and that the remnant has a shell of fast ejecta that interacts with interstellar or circumstellar gas. The latter is not yet confirmed for the Crab \citep[see, for example,][]{LunTzi12}. Whilst the Crab originates in an 8--10\,{\msun} star \citep[e.g.,][]{Hester08}, the progenitor of {\snr} probably has a mass of about 20\,{\msun} \citep[hereafter \paperone]{Williams10,Lun11}. \citet{Seraf04} compare {\psr} and the Crab. \cite{Williams08} review \psr, its wind nebula, as well as the entire supernova remnant (SNR).

Closer to the centre, {\snr} appears to be more asymmetric than the Crab Nebula, with much of the emission coming from a region a few arcseconds southwest of the pulsar. \citet[hereafter \pmorse; also see \citealt{Kirshner89} and \citealt{Seraf05}, herafter \paperser]{Morse06} show that this asymmetric appearance -- as well as a red-ward asymmetry in integrated emission-line profiles -- does not signal an overall asymmetry of the PWN, as the maximum velocity of emission lines towards and away from us are nearly equal. A detailed comparison between the PWNe of the two pulsars is more difficult due to the large distance of {\psr} \citep[about 51\,kpc,][]{Panagia05}, compared to the Crab \citep[$2\!\pm\!0.5$\,kpc; see][and references therein]{Kaplan08}; the semi-major axis of the torus in the Crab only subtends about 1\farcs5 at the distance of \psr. Because of this there is still no counterpart for {\psr} and its PWN to the detailed synchronised visual and X-ray study of the Crab PWN by \citet{Hester02}. The latter authors reveal the presence of outward moving equatorial wisps at half the lightspeed. Many time-variable sub-arcsecond structures exist in the Crab PWN in the visual wavelength range \citep{Hester02}, the near-infrared wavelength range \citep{Melatos05}, and in the X-ray regime \citep{Hester02,Weisskopf00}; this is all discussed in detail by \citet{Hester08}.

\citet{DeLuca07} are the first to report evidence for changes in the structure of {\snr} over a relatively short time period. They find a ten-year flux variation (1995--2005) in the southwest direction where the PWN emits the strongest, and suggest that they have detected a hot spot that moves at a speed of about four per cent of the lightspeed. They also note that the hot spot could be similar to a time-varying arc-like feature in the outer Crab Nebula, and that a pulsar jet in {\snr} could be directed towards the bright southwest region rather than perpendicular to this direction, as is earlier suggested by \citet{GW00}.

More recently {\paperone} present a study of the PWN of {\psr}; we use the same data as \citet{DeLuca07}, but complement it with all available Hubble Space Telescope (\textit{HST}) data. In particular, we include \textit{HST} polarisation data from 2007, as well as all available X-ray data. These data confirm a remarkable activity in the southwest region of the PWN, with the apparent ``blob'' 1\farcs3 southwest of the pulsar, which emerges in 1999 and then fades until 2005--2007. We suggest that the ``blob'' structure may be a part of now fading filamentary structure that was excited in 1999, rather than a moving entity. Particularly evident is how well the continuum emission traces the emission structure in [\ion{S}{ii}], and how low the correlation is between emission in the continuum and [\ion{O}{iii}]. A high intensity ratio $I_{\text{[\ion{S}{ii}]}}/I_{\text{[\ion{O}{iii}]}}$ can mean either shock activity \citep{Williams08}, or a locally high S/O-ratio. The former scenario could, e.g., be an outflow in the PWN from the pulsar, which temporally excites filaments at the ``blob''. {\paperone} discard time-dependent photoionization by emission from the X-ray blob as the source of the enhanced $I_{\text{[\ion{S}{ii}]}}/I_{\text{[\ion{O}{iii}]}}$, and instead argue for slower shocks that are transmitted into PWN-embedded filaments; in a way similar to the shocks that are transmitted into the blobs of the circumstellar ring around SN\,1987A \citep[see, for example,][]{Gron08}.

Unlike imaging observations, spectroscopic observations allow measurements of separate emission lines and kinematics. {\pmorse}, for example, present long-slit observations that are offset by 1{\arcsec} west of \psr, and measure various emission lines in the wavelength range 3335--6945\AA. They also present a one-dimensional kinematic structure along the slit (cf.\ their fig.~5). It is evident that a spatially resolved spectroscopic study in the visual wavelength range, which completely covers the central parts of {\snr} in several emission lines, would provide new information. The derived velocity field could be used to create a three-dimensional model of the PWN, which can be used to separate and analyse different structures in the object. Here we present and discuss such new observations of the central structure of {\snr}. For this purpose we used the fibre-fed integral-field spectrograph VIMOS at the Very Large Telescope (VLT) of the European Southern Observatory (ESO) in Chile. We organised the paper as follows. In Sections 2 and 3 we describe and discuss the observations and the data-reduction procedure. In Section~4 we present the results, and follow up with a discussion in Section~5. We close the paper with conclusions in Section~6.

\section{Observations}\label{sec:observations}
The object was observed in service mode during the period P86 with the visual multi-object spectrograph \citep[VIMOS,][]{LeFevre03} that is mounted on the Nasmyth focus of the UT3 8.2m telescope at the VLT in Chile. We used the integral field unit (IFU) with the smaller scale setting, 0\farcs33. The HR-blue and the HR-orange grisms were used to observe [\ion{O}{iii}]$\,\lambda5007$ and [\ion{S}{ii}]$\,\lambda\lambda6717,6731$. HR-blue (HR-orange) approximately covers the wavelength range 4150--6200{\AA} (5250--7400\AA) at a resolution of $R\!=\!2550$ ($R\!=\!2650$) and a dispersion of $0.51\,\AA\,\text{pixel}^{-1}$ ($0.6\,\AA\,\text{pixel}^{-1}$), or 31{\kms} at 5007{\AA} (27{\kms} at 6717{\AA}). Our object is faint, which is why we used the (nearly) longest exposure time possible in service-mode observations, 1800\,s; with shorter exposures readout noise dominates faint regions with the 0\farcs33 scale setting. We present our complete observations journal in Table~\ref{t21:arch}. For each science exposure the five columns provide: the observation date, the ESO-specific observation block id number, the seeing, the airmass, and the grism. The airmass and the seeing were within the ranges 1.40--1.45, and 0\farcs50--0\farcs80, respectively. In this study we only used the best-seeing exposures that are prefixed with a bullet; we intend to combine and use all data in a future study.

\begin{table}
\caption{The observations journal of the {\psr} field.}
\label{t21:arch}
\begin{center}
\begin{tabular}{lllccl}
\hline
\hline
          & \multicolumn{1}{c}{Date}  & \multicolumn{1}{c}{Obs.} & Seeing & Airmass & \multicolumn{1}{c}{Grism}  \\ 
          &            &   \multicolumn{1}{c}{Block id}  & [arcsec] &       &        \\[0.5ex]
\hline
          & 2010 Nov.~1  & 504960  & 0\farcs82 & 1.40 & HR-orange \\
          & 2010 Nov.~1  & 504961  & 0\farcs84 & 1.41 & HR-orange \\
          & 2010 Nov.~3  & 504967  & 0\farcs78 & 1.40 & HR-blue   \\
          & 2010 Nov.~7  & 504959a & 1\farcs10 & 1.41 & HR-orange \\
          & 2010 Nov.~28 & 504953  & 0\farcs63 & 1.42 & HR-orange \\
$\bullet$ & 2010 Nov.~28 & 504958  & 0\farcs56 & 1.47 & HR-orange \\
          & 2010 Nov.~28 & 504959b & 0\farcs60 & 1.40 & HR-orange \\
$\bullet$ & 2010 Nov.~29 & 504962  & 0\farcs62 & 1.40 & HR-blue   \\
          & 2010 Nov.~29 & 504963  & 0\farcs74 & 1.44 & HR-blue   \\
          & 2010 Dec.~1  & 504964  & 0\farcs75 & 1.41 & HR-blue   \\
          & 2010 Dec.~1  & 504965  & 0\farcs75 & 1.45 & HR-blue   \\
          & 2010 Dec.~1  & 504966a & 1\farcs00 & 1.53 & HR-blue   \\
          & 2010 Dec.~2  & 504966b & 0\farcs95 & 1.44 & HR-blue   \\[0.5ex]
\hline
\end{tabular}\\
\end{center}
\end{table}

\begin{figure}
\centering
\includegraphics[width=\columnwidth]{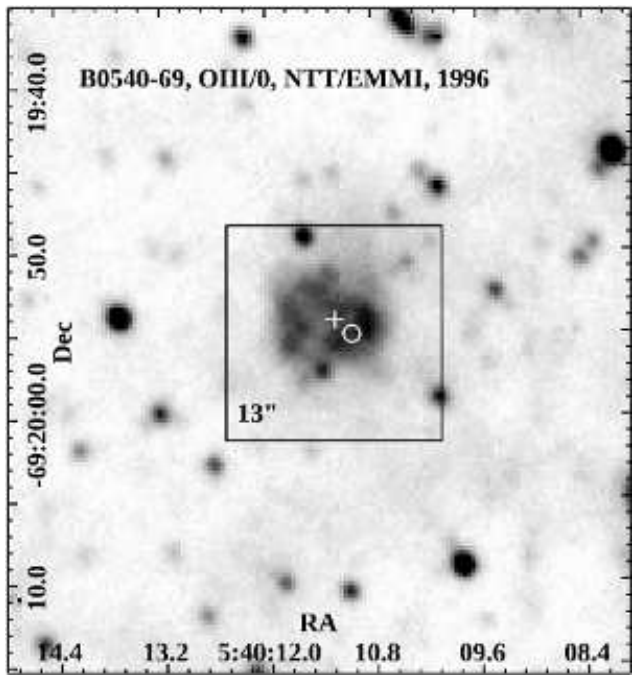}
\caption{An [\ion{O}{iii}]\,$\lambda5007$-filter $40\arcsec\!\times\!40\arcsec$ NTT/EMMI image from 1996 of the inner region of {\snr} (Lundqvist et al., in prep.). The 13\arcsec$\times$ 13{\arcsec} central part illustrates region that we observed with the VIMOS IFU using the 0\farcs33 scale setting. The pulsar (blob) position is marked by a white cross (circle).}
\label{fig:psr}
\end{figure}

In addition to the science exposures (mandatory) continuum-lamp and arc-lamp exposures were taken immediately after each science exposure, as well as spectrophotometric standard-star exposures for each grism and night. The normalised continuum lamp images were used to correct for varying fibre-to-fibre transmission. Separate observations of the sky were considered too costly. Altogether the four detectors of the VIMOS IFU hold $40\!\times\!40\!=\!1600$ separate fibres, where each fibre represents a spatial element, or a so-called spaxel, on the sky. In the 0\farcs33 sampling mode each pointing covers an area of $13\farcs2\arcsec\!\times\!13\farcs2\arcsec\!=\!174\arcsec^2$ on the sky. However, the outermost frame of spatial elements is vignetted, which is why the effective area coverage is $12\farcs5\times12\farcs5=157\arcsec^2$. Figure~\ref{fig:psr} shows the {\psr} region (Lundqvist et\,al., in preparation), as well as the coverage of the VIMOS IFU 0\farcs33 scale setting. This smaller setting clearly offers the best spatial coverage and resolution.

\section{Data reduction}\label{sec:datareduction}
Data were reduced using {\pthreed} \citep[revision 1506, which is part of version 2.1,][]{San10,San11,San12}\footnote{All papers are freely available at the \textsc{p3d} project web site \href{http://p3d.sf.net}{http://p3d.sf.net}.}. VIMOS data are provided in files of four separate detectors per field, where each detector holds 400 full spectra. The files of each detector are reduced separately, and are combined after flux calibration, to create an image containing the full field of all 1600 spectra. Here follows a description of the method.

At first the five provided bias images of each night are combined to create a master-bias image, which is subtracted from all other images of the same night. In a second step, a trace mask is created using a continuum lamp image. Since VIMOS spectra overlap significantly on the CCD, cross-dispersion line profiles are calculated in this step, to allow a consecutive spectrum deconvolution. Cross-dispersion line profiles are calculated every 20th pixel on the dispersion axis to allow an accurate extraction, they are interpolated for the remaining pixels. In a third step a dispersion mask for the wavelength calibration is created using an arc image. There are few and mostly no visible arc lines blue-wards of the helium line at 5852{\,\AA} in data of the HR-orange grism. This is no problem, however, since we are only interested in [\ion{S}{ii}]$\,\lambda\lambda6717,6731$. The residual $r$ between the fitted wavelength, which is calculated using a fifth order polynomial, and the known wavelengths of the arc lines, is always in the range $0.02\!\la\!r\!\la\!0.06\,\text{\AA}$.

Cosmic-ray hits are removed, before spectra are extracted, in the science images using the \textsc{PyCosmic} algorithm \citep{Husemann12}; in the version that is integrated into \pthreed, using default parameters. Furthermore, \citet{San10} show that accurate spectrum centre positions are crucial when extracting spectra accurately, even an offset of 0.1 pixels affects results significantly. With VIMOS calculated profiles have to be offset due to instrument flexure between the times of the continuum lamp exposure and the science exposure. New centre positions are calculated (for working fibres) in the science image, which is first median-filtered on the dispersion axis. One median value of the difference between all old and new centre positions is calculated and added to the profile positions. In a fourth step spectra are extracted using the multi-profile deconvolution method (\citealt{San10}, which is the \pthreed-version of the algorithm of \citealt{Sharp10}).

Data of all four quadrants of either grism are setup to use the same wavelength array. This choice allows a straightforward combination of reduced and flux-calibrated images into one spectrum image. The dispersion mask was then applied in a fifth step. In a sixth step a correction to the fibre-to-fibre sensitivity variations was applied to the extracted and wavelength-calibrated science spectra by dividing with the mean spectrum of an extracted and normalised continuum lamp flat-field image, excluding all spectra of broken and low-transmission fibres. CCD fringing effects are removed rather well in this step as the flat-field spectra are neither smoothed nor replaced with a polynomial, which is the usual procedure with {\pthreed} for other instruments. During extraction data are changed from a CCD-based format to a row-stacked-spectra (RSS) format.

In a seventh step, extracted spectra are flux calibrated using the reduced standard-star exposure of the respective night and grism. In an eighth step flux calibrated images of the four separate detectors are combined to create one RSS image with 1600 spectra. For unknown reasons, as well as insufficient information to do calibration even more accurately, measured fluxes vary with both the detector and with spatial elements \citep[also see][]{Lag12}. All spectra are therefore normalised to achieve the same integrated median flux in a telluric line. We used the telluric line near $\lambda\!=\!4866\AA$ ($\lambda\!=\!6864\AA$) with data of HR-blue (HR-orange). In step nine, the detector-combined images are corrected for differential atmospheric refraction (DAR) as is described in \citet{San12}. Finally, in step ten extracted RSS images are converted to data cubes, where the first two dimensions contain spatial information -- East is to the left and North is up -- and the third dimension contains the spectrum information.

\section{Results}
To extract velocity structures of gases emitting in spectral lines that are available in our data, we focus on [\ion{S}{ii}]\,$\lambda\lambda6716,6731$ and [\ion{O}{iii}]\,$\lambda\lambda4959,5007$. These two multiplets produce different structures in {\sl HST} imaging observations (\paperone). In particular, we find a spatial anti-correlation between [\ion{S}{ii}] and [\ion{O}{iii}] at the position of a ``blob'' about 1\farcs3 southwest of the pulsar, with [\ion{S}{ii}] tracing the optical continuum emission, as well as the surface intensity of X-rays. 

Potential problems with imaging observations are contamination by point sources -- including stars and the pulsar -- and the LMC background, also the red wings of [\ion{S}{ii}]\,$\lambda6716$ may blend with the blue wing of [\ion{S}{ii}]\,$\lambda6731$, and only a fraction of the combined multiplet flux enters a spectral filter. The difference in wavelength between the two sulphur (oxygen) lines corresponds to $641\kms$ ($2890\kms$). In comparison, the total line width of [\ion{O}{iii}] is at least $3300\kms$ (\pmorse), which means that both lines have blended line components.

With our spectroscopic VIMOS data we are able to bypass several of these problems. At first, we note that out of 1600 spectra in one science exposure about 200 passed through broken fibres, are vignetted, or are otherwise affected by such broken and vignetted spectra during the DAR correction. Hence, in our analysis we used the (approximately) 1400 remaining spectra. Moreover, for each individual useful spectrum we analysed the full [\ion{S}{ii}] and [\ion{O}{iii}] multiplets. We made linear fits to the continuum levels within the spectral domains of each multiplet, and then subtracted these continua from the respective spectra. We did not deredden the spectra.

In the following we first discuss the [\ion{O}{iii}] and the [\ion{S}{ii}] spectra in Sections 4.1 and 4.2. Thereafter, we study intensity and density maps in Section 4.3, and velocity maps in Sections 4.4 and 4.5.

\subsection{Details of the [\ion{O}{iii}] spectrum}
For [\ion{O}{iii}] we decomposed the two line components, using the knowledge that the transition probability ratio $A(\text{[\ion{O}{iii}]}\,\lambda5007)/A([\text{\ion{O}{iii}]}\,\lambda4959)\!=\!3$. To remove the LMC back\-ground, we fitted the spectrally unresolved Gaussian line profiles in areas near the four corners of the field (avoiding spectra of the low-transmission fibres in the second quadrant). Thereafter, we created a two-dimensional fit to subtract that LMC component from all spectra. Simultaneously we transferred the spectrum to the velocity of the LMC [\ion{O}{iii}] emission in the field, which we found to be $v_{\text{LMC}}=+286.5\kms$.

\begin{figure}
\centering
\includegraphics[width=85mm, clip]{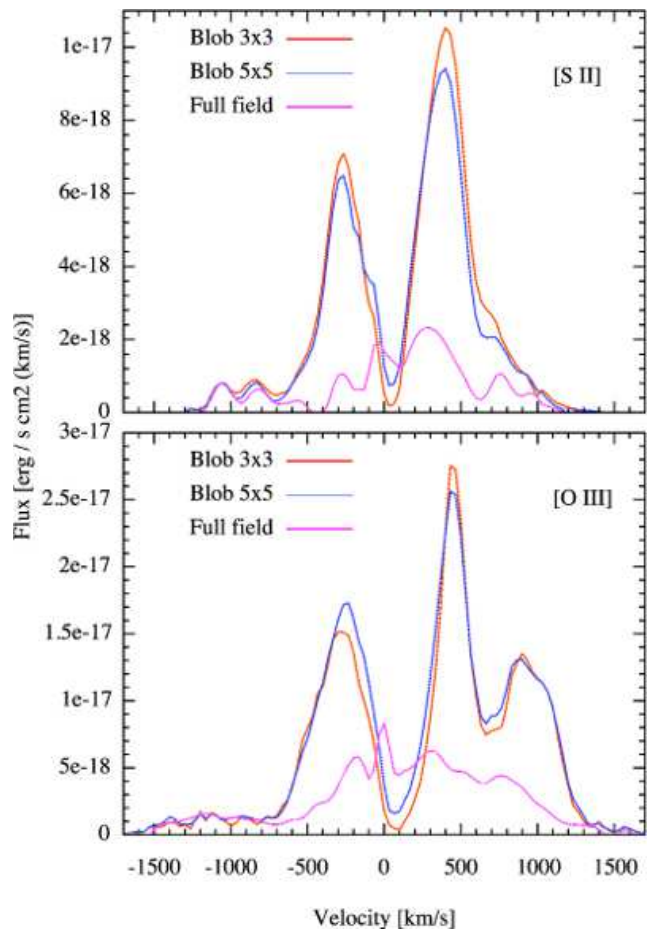}
\caption{Average spectra of [\ion{S}{ii}]\,$\lambda\lambda6716,6731$ (top panel) and [\ion{O}{iii}]\,$\lambda5007$ (bottom panel) for the entire field (magenta; which includes about 1400 fibres out of $40\!\times\!40$ available fibres) and for a region that is centred on the ``blob'' that is discussed in \paperone. The red (blue) line shows the average spectrum of $3\!\times\!3$ fibres ($5\!\times\!5$ fibres). The average profiles for [\ion{S}{ii}] were drawn using a ``master'' [\ion{S}{ii}] line profile, which contains information of both [\ion{S}{ii}]\,$\lambda6716$ and [\ion{S}{ii}]\,$\lambda6731$.}
\label{fig:blobspec}
\end{figure}

The average spectrum of all 1400 fibres in the entire field for [\ion{O}{iii}]\,$\lambda5007$ is shown with a magenta-coloured line in the lower panel of Fig.~\ref{fig:blobspec}. The full line profile fills the velocity range \mbox{-$1650\le v_{\text{\ion{O}{iii}}}\le+1700\kms$}, i.e., we recover the total width that is observed by \pmorse; we note that our spectral resolution is about eight times as high as theirs (250\kms); their line profile could therefore be slightly broader for the same area probed. (\citealt{LunTzi12} discuss the importance of sufficient spectral resolution to constrain the faint parts of line profiles in the Crab SNR.) Apart from that, we cannot obtain a perfect match between our data and those of \pmorse, despite their wide (2\farcs0) slit and fairly poor (1\farcs5) seeing, which should include much of the PWN in the detected emission. The most obvious features are, however, identical, i.e., the generally redshifted emission with a broad shoulder in the blue. What may be more pronounced in our data is the stronger shoulder emission in the blue at velocities $v_{\text{\ion{O}{iii}}}\lsim$-$1300\kms$ than red-wards of $v_{\text{\ion{O}{iii}}}\gsim\!+1300\kms$.

Our VIMOS data allow a detailed study of spectra for smaller regions of the PWN. This is highlighted by the blue and the red spectra in the lower panel of Fig.~\ref{fig:blobspec}. The spectrum that is drawn in red colour is the average spectrum of $3\!\times\!3$ fibres (corresponding to a $1\arcsec\!\times\!1\arcsec$ area), which encapsulates the ``blob'' in \paperone. The spectrum that is drawn in blue colour is the average spectrum of $5\!\times\!5$ fibres, which are also centred on the ``blob''. In general, the average spectra of the ``blob" are much stronger than the average spectrum of the entire field, and there is nearly no emission near $v_{\text{\ion{O}{iii}}}=+100\kms$. This low emission is flanked by peaks at about $v_{\text{\ion{O}{iii}}}\!=$-$250\kms$ and $v_{\text{\ion{O}{iii}}}\!=\!+450\kms$, as well as a second peak at $v_{\text{\ion{O}{iii}}}\!=\!+900\kms$. Compared to the 1400-spectrum average line profile, the ``blob'' spectrum is centred to the red, with a possibly higher red shoulder at $v_{\text{\ion{O}{iii}}}\!\gsim\!+\!1300\kms$. The blue shoulder at velocities $v_{\text{\ion{O}{iii}}}\!\lsim$-$1200\!\pm\!300\kms$ is very similar to that of the spectrum of the entire field. This suggests that the blue shoulder is a general feature that exists across the entire field. We discuss this further in Section 4.4.

\subsection{Details of the combined [\ion{S}{ii}] spectrum}
The LMC background is lower at the emission lines of [\ion{S}{ii}] than at [\ion{O}{iii}], but a complication with recovering a clean [\ion{S}{ii}] line profile is the line ratio $c_{\text{[\ion{S}{ii}}]}\!=\!I($[\ion{S}{ii}]\,$\lambda6731)/I($[\ion{S}{ii}]\,$\lambda6716)$, which depends on the electron density. {\paperser} identify a region where the two line components do not overlap, and use this region to derive a line ratio and then the electron density $\eden\!=\!(1-5)\times10^3$~cm$^{-3}$. Here we decomposed the two line components for all spectra where the signal-to-noise is high enough. Next we describe the procedure.

\begin{figure*}
\centering
\includegraphics[width=180mm, clip]{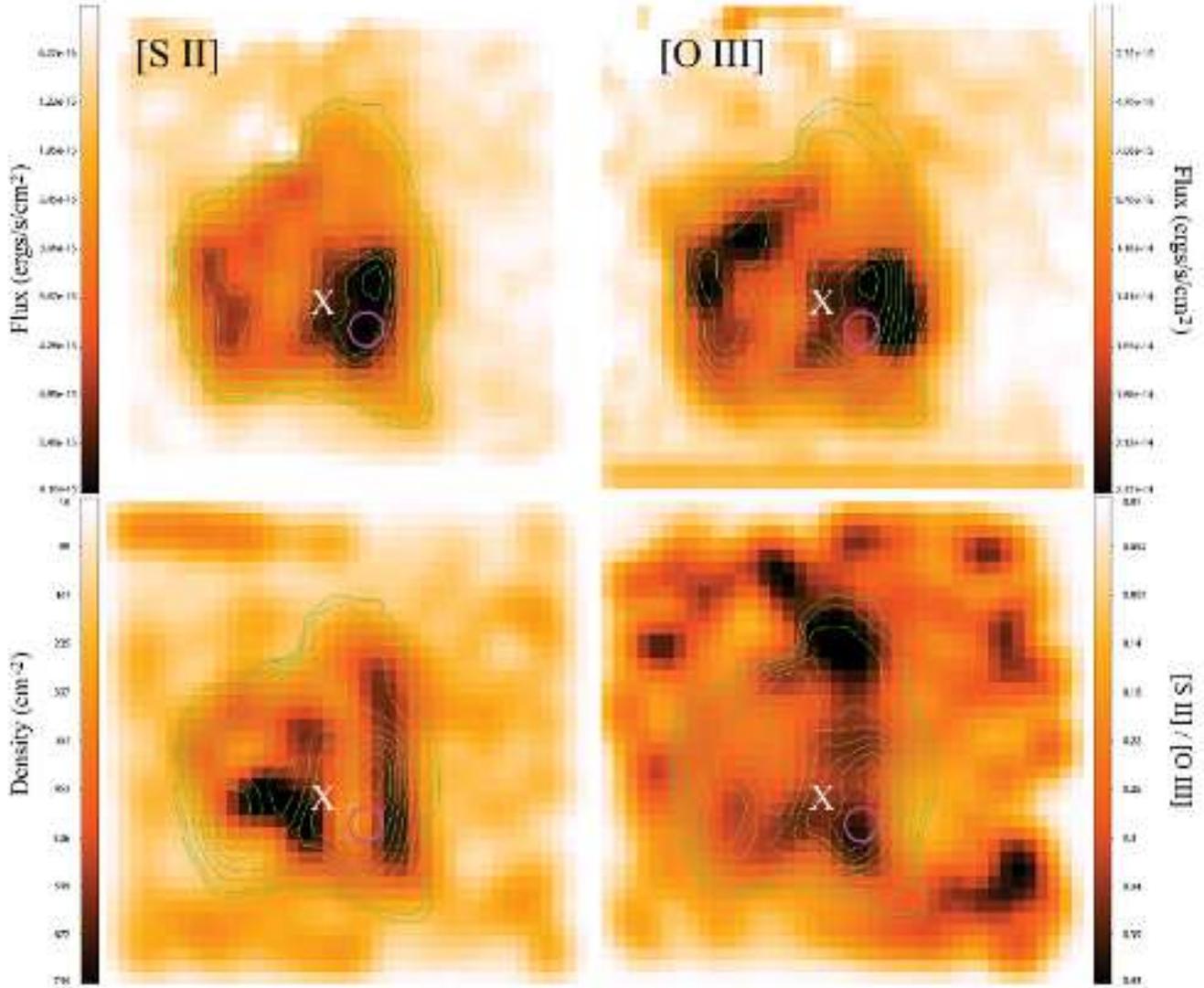}
\caption{For the innermost $11\farcs22 \times 11\farcs22$ ($2.77\!\times\!2.77\,\text{pc}^2$) area of {\snr}, the flux of [\ion{S}{ii}] ([\ion{O}{iii}]) is shown in the upper left-hand side (right-hand side) panel. The intensity unit is intensity per 0\farcs33~$\times$~0\farcs33 pixel. The images show no continuum sources, since we subtracted the continuum for each spatial element individually. Data were smoothed in a convolution with a $2\!\times\!2$-pixel kernel. The lower left-hand side (right-hand side) panel shows the electron density {\eden} ($I_{\text{[\ion{S}{ii}]}}/I_{\text{[\ion{O}{iii}]}}$). North is up and east to the left. Each spatial element is $0\farcs33\!\times\!0\farcs33$. Spatial elements around the border are differently affected by effects of the DAR correction. The spatial elements in the northeast region show emission that passed through low-transmission fibres in the second quadrant. The contours of the [\ion{S}{ii}] flux are, moreover, reproduced in all panels. The position and size of the ``blob'' (\paperone) is indicated by a magenta-coloured circle and the pulsar position is marked by a white X.}
\label{fig:full_images}
\end{figure*}

After we removed the continuum emission in the same way as for [\ion{O}{iii}], we fitted the line spectrum (before decomposition) using a Savitzky-Golay filter to mimic noise influence. We then swept from the blue side using [\ion{S}{ii}]\,$\lambda6716$ as a template, and from the red side using  [\ion{S}{ii}]\,$\lambda6731$ as a template. From $c_{\text{[\ion{S}{ii}}]}$ we constructed a ``master'' [\ion{S}{ii}] line profile $f_\lambda$:
\begin{eqnarray*}
f_\lambda = wf_{\lambda 1} + \left(1-w\right) f_{\lambda 2},\quad w = \frac{1}{1+c_{\text{[\ion{S}{ii}}]}},
\end{eqnarray*}
where $f_{\lambda 1}$ is the forward sweep for [\ion{S}{ii}]\,$\lambda6716$ and $f_{\lambda 2}$ the backward sweep for [\ion{S}{ii}]\,$\lambda6731$. For low-density gas dominating the spectrum, [\ion{S}{ii}]\,$\lambda6716$ dominates the ``master'' [\ion{S}{ii}] line profile, and for high-density gas [\ion{S}{ii}]\,$\lambda6731$ dominates. We calculated $c_{\text{[\ion{S}{ii}}]}$ to give the best correlation between the blue-ward sweep using [\ion{S}{ii}]\,$\lambda6716$ as template, and the red-ward sweep using [\ion{S}{ii}]\,$\lambda6731$ as template. The average correlation coefficient between those functions for the central $34\!\times\!34$ fibres is 0.90. For the $5\!\times\!5$ fibres for the ``blob'' in Fig.~\ref{fig:blobspec}, the average correlation coefficient is 0.95.

\begin{figure*}
\centering
\includegraphics[width=160mm, clip]{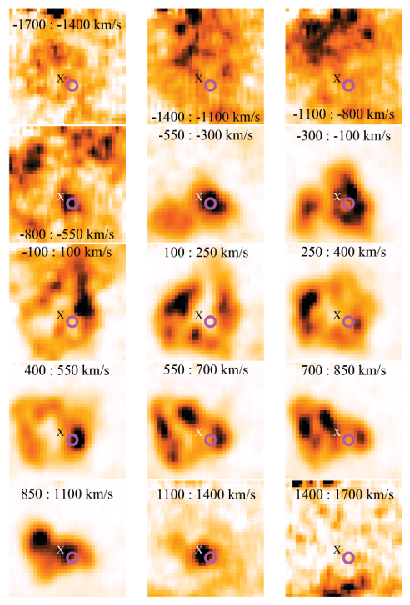}
\caption{Intensity maps of [\ion{O}{iii}] at separate and specified velocity intervals for the innermost $10\farcs56\!\times\!10\farcs56$ ($2.61\!\times\!2.61\,\text{pc}^2$) area of \snr, each spatial element is $0\farcs33\!\times\!0\farcs33$. The position and size of the ``blob'' (\paperone) is indicated by a magenta-coloured circle and the pulsar position is marked by an X. Colours are individual to each panel to emphasize structures in the selected velocity range.}
\label{fig:o3vel}
\end{figure*}

To remove the LMC background, we used a similar procedure as with [\ion{O}{iii}]. The velocity of the LMC [\ion{S}{ii}] emission, was also in this case found to be $v_{\text{LMC}}\!=\!+286.5\kms$. The average of all 1400 spectra is shown in the upper panel of Fig.~\ref{fig:blobspec} (magenta). The full [\ion{S}{ii}] line profile is not as wide as the corresponding line profile of [\ion{O}{iii}]\,$\lambda5007$, \mbox{-$1200\!\le\!v_{\text{[\ion{S}{ii}]}}\!\le\!1200\kms$}. Compared to [\ion{O}{iii}], the [\ion{S}{ii}] line profile is more heavily biased towards redder wavelengths, with a pronounced peak at $v_{\text{[\ion{S}{ii}]}}\!\approx\!+300\kms$. As with [\ion{O}{iii}], there could be a shoulder in the blue, but this only extends out to $v_{\text{[\ion{S}{ii}]}}\!=$-$1200\kms$.

The blue and the red spectra in the upper panel of Fig.~\ref{fig:blobspec} highlight the area around the ``blob''. The difference in emission per spectrum between the ``blob'' region and the entire field is even more pronounced for [\ion{S}{ii}] than for [\ion{O}{iii}]. There is also here nearly no emission for $0\!\le\!v_{\text{[\ion{S}{ii}]}}\!\le\!100\kms$, and there are peaks at $v_{\text{[\ion{S}{ii}]}}\!\approx$-$300\kms$ and $v_{\text{[\ion{S}{ii}]}}\!\approx\!+400\kms$; although, unlike for [\ion{O}{iii}] there is no peak at $v_{\text{[\ion{S}{ii}]}}\!\approx\!+900\kms$.

\begin{figure*}
\centering
\includegraphics[width=180mm, clip]{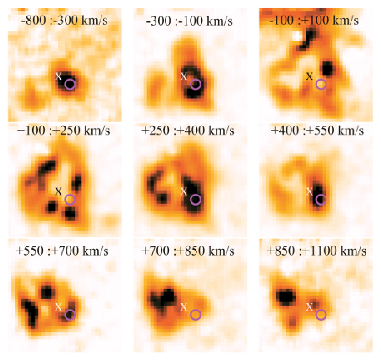}
\caption{Intensity maps of [\ion{S}{ii}] at separate and specified velocity intervals for the innermost $10\farcs56\!\times\!10\farcs56$ ($2.61\!\times\!2.61\,\text{pc}^2$) area of \snr, each spatial element is $0\farcs33\!\times\!0\farcs33$. For more details see the caption of Fig.~\ref{fig:o3vel}.}
\label{fig:s2vel}
\end{figure*}

\subsection{Intensity and density maps of both [\ion{O}{iii}] and [\ion{S}{ii}]}
We integrated individual spectra of [\ion{O}{iii}] and [\ion{S}{ii}] in velocity to generate projected intensity maps, see the upper two panels of Fig.~\ref{fig:full_images}. To highlight differences in the surface-intensity distribution of [\ion{O}{iii}] and [\ion{S}{ii}], we show the intensity ratio $I_{\text{[\ion{S}{ii}]}}/I_{\text{[\ion{O}{iii}]}}$ in the lower right-hand side panel; again, note that the spectra were not extinction corrected. We also show an electron-density (\eden) map in the lower left-hand side panel, which is derived from the decomposition of the two line components of [\ion{S}{ii}]. To calculate the electron density, we used our multilevel model atom that is described in \citet{Mar00}.

A few regions stand out in a comparison of the maps of [\ion{O}{iii}] and [\ion{S}{ii}]. To the northeast, there is a region where both $I_{\text{[\ion{S}{ii}]}}/I_{\text{[\ion{O}{iii}]}}$ and {\eden} are low. A similar region exists immediately west of the ``blob'', where {\eden} is higher. The intensity ratio $I_{\text{[\ion{S}{ii}]}}/I_{\text{[\ion{O}{iii}]}}$ is high for the ``blob'' itself, and in the northwest region of the PWN. Note the low density of the ``blob''. Moreover, high densities are seen in the southeast, but neither [\ion{O}{iii}] nor [\ion{S}{ii}] are particularly bright there. In general, we do not recover the high value of $\eden\!=\!(1-5)\times10^3$~cm$^{-3}$ of {\paperser}. Instead, the electron density in the [\ion{S}{ii}] filaments is $\la\!10^3\,\text{cm}^{-3}$, the highest value is about $750\,\text{cm}^{-3}$ $<\!1\arcsec$ west of the blob and 1--$3\arcsec$ east of the blob. We caution, however, that the maps in Fig.~\ref{fig:full_images} were smoothed over two pixels, which results in smeared out density peaks. In unsmoothed data there are single spatial elements where $\eden\!\simeq\!2\!\times\!10^3\,\text{cm}^{-3}$, and in this sense the high value of {\paperser} is recovered; their value is measured at one position along their slit where the two components of [\ion{S}{ii}] do not blend.

In a comparison with the wavelet-filtered \textit{HST} images in fig.~13 of \paperone, we recover the general structures of the \textit{HST} [\ion{S}{ii}] and [\ion{O}{iii}] images well. The difference in epoch between the \textit{HST} and the VLT data sets is about eleven years. A peak in $I_{\text{[\ion{S}{ii}]}}/I_{\text{[\ion{O}{iii}]}}\!\simeq\!0.4$ occurs at the ``blob'' position in both data sets. Another peak with the same value on the ratio is seen in the northwest region, where the remnant bulges out in [\ion{S}{ii}] (cf.\ Fig.~\ref{fig:full_images}); an enhanced ratio is also present at that position in the \textit{HST} intensity-ratio plot.

\subsection{Intensity maps at separate velocities}
Our spectroscopic data contain more information than the projected images in Fig.~\ref{fig:full_images} reveal. We show intensity maps of separate velocity intervals for [\ion{O}{iii}] in Fig.~\ref{fig:o3vel} and for [\ion{S}{ii}] in Fig.~\ref{fig:s2vel}. These figures are smoothed in a convolution using a $3\!\times\!3$-pixel kernel. We only show the central $32\!\times\!32$ spatial elements ($10\farcs56\!\times\!10\farcs56$ or $2.61\!\times\!2.61\,\text{pc}^2$, assuming LMC is 51\,kpc away, cf.\ Section~\ref{sec:threed}).

\begin{figure*}
\centering
\includegraphics[width=160mm]{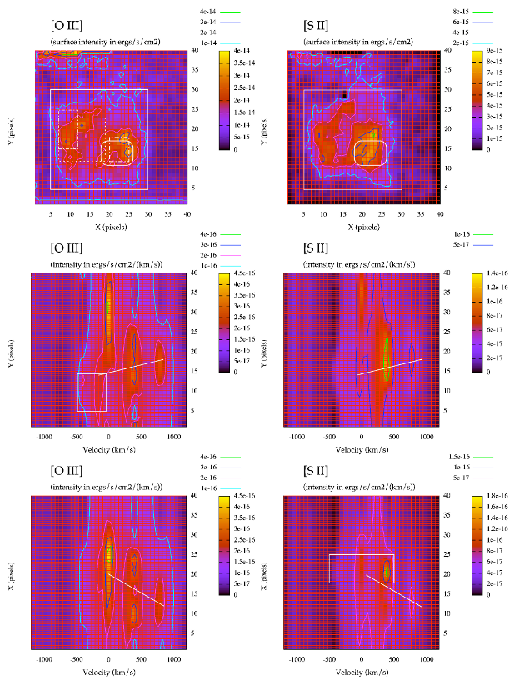}
\caption{Integrated data of three-dimensional cubes in [\ion{O}{iii}] (left-hand side column) and [\ion{S}{ii}] (right-hand side column) as they appear on the sky (upper row). Increasing values on the $x$-axis (y-axis) go from East to West (South to North). The second (third) row is obtained by summing values along the East-West (South-North) direction. White lines and boxes indicate regions that are discussed in the text. A raster has been put on the panels for orientation; one cell is 62.5{\kms} wide on the velocity axes. The ``blob'' is centred on $(x,y)\!\approx\!(22.1,13.7)$.}
\label{fig:cubes}
\end{figure*}

We begin with gas that moves the fastest towards us -- the three panels in the uppermost row of Fig.~\ref{fig:o3vel} for [\ion{O}{iii}]. There is an extended area of emission, with some enhancement, in the northeast. This emission strengthens our hypothesis in Section 4.1 that the blue shoulder of the 1400-spectrum and the ``blob'' spectrum in Fig.~\ref{fig:blobspec} possibly is an extended fast shell-like structure in [\ion{O}{iii}] on the near side of the remnant, as seen from the Earth. A corresponding emission, or ``haze'', is visible at high positive velocities (Fig.~\ref{fig:mayavi2}), whilst there is no clear red shoulder in Fig.~\ref{fig:blobspec}. The ``haze" is thus weaker on the red side than on the blue side.

The ``blob'' emerges in the velocity interval -$800\!\le\!v\!\le$\mbox{-300}\kms, in both [\ion{O}{iii}] and [\ion{S}{ii}]. A pronounced region southeast of the ``blob'' is also visible. For -$300\!\le\!v\!\le$-$100\kms$, the ``blob'' becomes encircled by a shell that breaks up at higher (redder) velocities; the ``blob'' appears to move at a lower velocity. A point-like structure re-emerges at the ``blob'' position for $400\!\le\!v\!\le\!550\kms$; this explains the line profiles in Fig.~\ref{fig:blobspec}. The main difference between [\ion{O}{iii}] and [\ion{S}{ii}] is, furthermore, that [\ion{S}{ii}] is also very pronounced for $250\!\le\!v\!\le\!400\kms$, whereas [\ion{O}{iii}] peaks immediately west of the ``blob''. Figure~\ref{fig:o3vel} shows [\ion{O}{iii}] emission from the ``blob'' for $850\!\le\!v\!\le\!1400\kms$, which confirms the high-velocity peak in [\ion{O}{iii}] in Fig.~\ref{fig:blobspec}. This structure also appears to be connected to a region northeast of the ``blob''. For the ``blob'' it seems as if it could be a separate entity for -$800\!\le\!v\!\le\!700\kms$. The high value of $I_{\text{[\ion{S}{ii}]}}/I_{\text{[\ion{O}{iii}]}}$ for the ``blob'' in Fig.~\ref{fig:full_images} is likely due to strong [\ion{S}{ii}] emission for $250\!\le\!v\!\le\!400\kms$, where [\ion{O}{iii}] is fainter.

Earlier in this section we highlighted additional regions. The region to the northeast, with a low value on both $I_{\text{[\ion{S}{ii}]}}/I_{\text{[\ion{O}{iii}]}}$ and {\eden}, appears to be dominated by [\ion{O}{iii}] for $100\!\le\!v\!\le\!250\kms$. The (high-density) region immediately west of the blob, which shows a low value on $I_{\text{[\ion{S}{ii}]}}/I_{\text{[\ion{O}{iii}]}}$, emits strongly in [\ion{O}{iii}] at many velocities in the range \mbox{-$300\!\le\!v\!\le\!+1100\kms$}. In the same region, the west streak of high electron density and high $I_{\text{[\ion{S}{ii}]}}/I_{\text{[\ion{O}{iii}]}}$ in Fig.~\ref{fig:full_images} -- that extends from north to south between the protrusions in [\ion{S}{ii}] to the northeast and the southeast -- is marked mainly for -$100\!\le\!v\!\le\!250\kms$. Despite a high density, this feature could partly be due to poorly subtracted LMC emission.

\subsection{A closer study of the velocity structures}
In Fig.~\ref{fig:cubes} we present projected intensities of the three-dimensional data cube of [\ion{O}{iii}] and [\ion{S}{ii}]. The intensities were summed along the line of sight (the $x$-axis; the $y$-axis) in the upper (middle; lower) panel of the figure, where the $x$-axis (velocity-axis; velocity-axis) extends rightwards from East to West (approaching-to-receding; approaching-to-receding) and the $y$-axis ($y$-axis; $x$-axis) upwards from South to North (South to North; East to West). The two upper panels show the lines as they appear on the sky. To guide the eye we put a raster on top of each image. The ``blob'' is centred at $(x,y)\approx(22.1,13.7)$. The images were not smoothed, unlike the data in Fig.~\ref{fig:full_images}. To see details in the central part of the remnant, we focused on the velocity interval \mbox{-$1200\!\leq\!v\!\leq\!1200\kms$.} The absolute intensity-scale panels in Fig.~\ref{fig:cubes} complement those in Figs.~\ref{fig:o3vel} and \ref{fig:s2vel}, since the panels in the latter two figures use different intensity cuts to highlight various features in the individual panels.

In the second row of Fig.~\ref{fig:cubes}, we clearly see three distinct velocity features: one near the LMC rest velocity, another broader for $200\!\le\!v\!\le\!500\kms$, and a third one near $v\!=\!800\kms$. The first feature is located to the northern part of the data cube and is most likely influenced by the LMC background. Focusing on the structures between $(5,5)\!\leq\!(x,y)\!\leq\!(30,30)$ (the region is indicated with a large white box in both top-row panels), it is clear that some of the emission about zero velocity at higher $x$ values may not belong to the remnant. The situation is different for $5\!\leq\!y\!\leq\!15$, where there is a blueshifted feature in [\ion{O}{iii}] at about $v\!=$-$200\kms$, which also continues out to $v\!\approx$-$500\kms$ (indicated with a white box in the middle left-hand side panel). For both [\ion{O}{iii}] and [\ion{S}{ii}], there is enhanced emission about the $y$-value for the ``blob'', i.e., $(18,11)\!\leq\!(x,y)\!\leq\!(25,17)$ (the ``blob'' is indicated with a white box with rounded corners in both top-row panels; this is more easily seen in the on-line version). Furthermore, in the third row of Fig.~\ref{fig:cubes} the ``blob'' region clearly sticks out; [\ion{S}{ii}] emission from the ``blob'' region begins at about $v\!=$-$500\kms$ and ends with bright emission at about $v\!=\!400\kms$ (indicated with a white box). This is also seen in the velocity plots Figs.~\ref{fig:o3vel} and \ref{fig:s2vel}, where the ``blob'' is clearly seen for -$800\!\le\!v\!\le$-$300\kms$.

The second velocity component in the second row of panels, $200\!\le\!v\!\le\!500\kms$, subtends the interval $(7,8)\!\le\!(x,y)\!\le\!(27,27)$ (using the third-row panels; the region falls just inside the large white box in the top panel). In a comparison with Figs.~\ref{fig:o3vel} and \ref{fig:s2vel}, we clearly see that there are mainly two emitting regions for $200\!\le\!v\!\le\!500\kms$. In [\ion{O}{iii}] the two regions are $(20,12)\!\le\!(x,y)\!\le\!(25,18)$ and $(7,15)\!\le\!(x,y)\!\le\!(12,25)$ (these two regions are indicated with dash-dotted boxes). In [\ion{S}{ii}], the former region dominates; this is clearly seen in the lower right-hand side panel of Fig.~\ref{fig:cubes}. Finally, the third component, at about 800\kms, is fainter and mainly emits in [\ion{O}{iii}]; it occupies the interval $(7,12)\!\le\!(x,y)\!\le\!(19,23)$ (indicated with a dotted box). This component is not the strongest at the ``blob'' position, but in the eastern region of the PWN.

An interesting feature of Fig.~\ref{fig:cubes} is the diagonal ubiquitous structure that extends in [\ion{O}{iii}] from $x\!\simeq\!20$ to $x\!\simeq\!12$, between $v\!\approx\!0\kms$ and $v\!\approx\!800\kms$, respectively, for the three major velocity features in the lower left-hand side panel (indicated with a white solid line). In the middle left-hand side panel, the diagonal is less obvious, but if we neglect the northern part of the velocity structure at the LMC rest velocity, we can draw a diagonal from $y\!\simeq\!14$ to $y\!\simeq\!18$ between \mbox{$v\!\simeq$-200\kms} and $v\!\simeq\!800\kms$ (indicated with a white solid line). The same diagonal structures also appear for [\ion{S}{ii}]. The ``blob'' is projected at the low-velocity ends of those two diagonals. Also note the symmetric features about those endpoints in velocity, for $v\!\simeq\!\pm400\kms$.

\begin{figure*}
\centering
\includegraphics[width=170mm, clip]{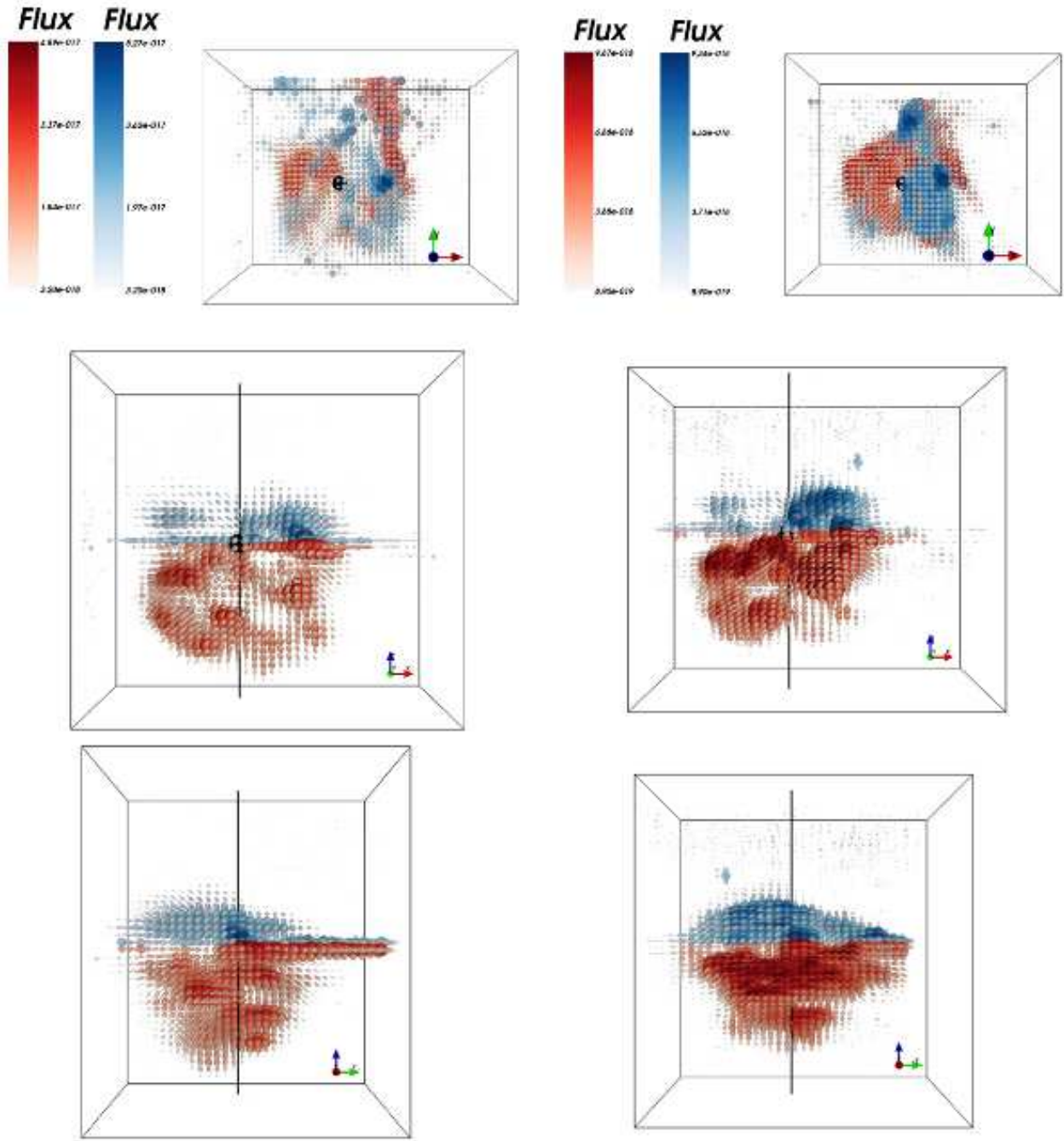}
\caption{Three-dimensional scenes of the central region of {\snr} that were created using the visualisation program \textsc{MayaVi}. The flux at each point is represented by both size and colour of partly transparent spheres. Zero velocity is at the local LMC rest velocity ($v_\text{LMC}\!=\!+286.5 \kms$). Blue colours indicate approaching gas at negative velocities -- i.e.\ positive $z$ -- and red colours indicate receding gas at positive velocities. Both colour bars are shown on the left-hand side of respective top panel. The three panels in the left-hand side (right-hand side) column show [\ion{O}{iii}] ([\ion{S}{ii}]) from different angles. The upper panels show the data cube as it appears on the sky; the $x$-axis ($y$-axis) extends rightwards from East to West (upwards from South to North), and the $z$-axis towards the Earth for decreasing velocities. The middle panels show the data cube as viewed in the plane of the sky from South to North inwards on the $y$-axis; the $x$-axis ($z$-axis) extends rightwards from East to West (upwards from longer wavelengths to shorter wavelengths). The lower panels show the data cube as viewed in the plane of the sky from East to West inwards on the $x$-axis; the $y$-axis ($z$-axis) extends rightwards from North to South (upwards from longer wavelengths to shorter wavelengths). The pulsar is identified with a black sphere in all panels; it was placed at the LMC rest velocity, using the exact optical coordinates of \psr. The vertical black line strikes through the pulsar and is parallel to the line of sight. The true position of the pulsar is somewhere along this line. The ``blob'' is located to the region that is void of ``blue'' emission in [\ion{O}{iii}], just southwest of the pulsar (top left-hand side panel), and overlaps with the region of ``blue'' emission in [\ion{S}{ii}] (top right-hand side panel); also compare with the projected intensity plots in Fig.~\ref{fig:full_images}. The axes of the two upper panels are contracted to leave space for the intensity bars.}
\label{fig:mayavi}
\end{figure*}

\section{Discussion}
In the following we model three-dimensional structures of {\snr} in Section 5.1, to interpret the new data. Thereafter, in Section~5.2 we compare our observations with previous studies.

\subsection{A closer study of the three-dimensional structures}\label{sec:threed}
In the previous section we found hints for diagonal structures, which originate in major features in the remnant. We speculate that such diagonal structures are part of bulk motions of the expanding gas. To explore this possibility further we used information of these ejecta to build a three-dimensional model of the remnant, assuming a linear expansion. Additional assumptions were an age of 1000 years and a distance of 51\,kpc \citep{Panagia05}. This implies that each $0\farcs33\!\times\!0\farcs33$ spatial element on the sky corresponds to $0.082\!\times\!0.082\,\text{pc}^2$, and that 1000\kms along the line of sight is 1.02\,pc. Because the seeing was about 0\farcs6, the spatial resolution is approximately 0.15\,pc. With a velocity resolution of 31\kms (cf.\ Section 2), the resolution along the line of sight is 0.032\,pc, i.e.\ about 4.6 times as good as the spatial resolution, and only a factor 2.8 worse than the best spatial resolution of \textit{HST}/WFPC2, 0.011\,pc, obtained with the Planetary Camera chip used in \paperone. We defined the $x$ and the $y$ axes as in Fig.~\ref{fig:cubes}, and added a $z$ axis that is oriented towards the observer, along the line of sight. We emphasise that the real structure along the $z$-axis -- as derived from motions along the line of sight -- may not be the same as the one we derive. For example, if gas emitting [\ion{O}{iii}] or [\ion{S}{ii}] was accelerated throughout the evolution or suddenly, as discussed in \paperone, structures may be exaggerated. However, if we focus on the larger features of the remnant, we could be close to the real structure. As another example, filaments in the Crab expand nearly linearly, and a linear expansion assumption is correct to within about 10\% \citep{Tri68}. The main uncertainty in a transformation from radial velocity to a physical scale for {\snr} is probably the uncertainty of the remnant age.

\begin{figure*}
\centering
\includegraphics[width=170mm, clip]{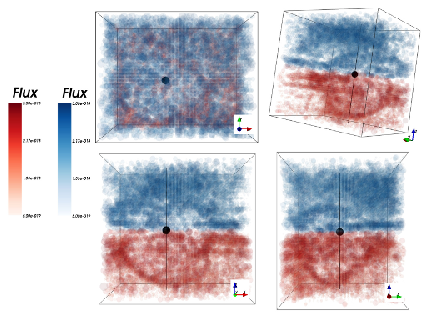}
\caption{Three-dimensional scenes of the [\ion{O}{iii}] faint emission in the central region of {\snr} that was visualised and presented in the same way as in Fig.~\ref{fig:mayavi}. The panels represent different orientations of the data cube. The upper left-hand side panel shows the data cube as it appears on the sky; the $x$-axis ($y$-axis) extends rightwards from East to West (upwards from South to North), and the $z$-axis towards the Earth for decreasing velocities. The upper right-hand side panel highlights the structure of faint emission. The lower left-hand side panel shows the data cube as viewed in the plane of the sky from South to North inwards on the $y$-axis; the $x$-axis ($z$-axis) extends rightwards from East to West (upwards from longer wavelengths to shorter wavelengths). The lower right-hand side panel shows the data cube as viewed in the plane of the sky from East to West inwards on the $x$-axis; the $y$-axis ($z$-axis) extends rightwards from North to South (upwards from longer wavelengths to shorter wavelengths). The pulsar is identified with a black sphere in all panels. The vertical black line strikes through the pulsar and is parallel to the line of sight.}
\label{fig:mayavi2}
\end{figure*}

With our derived properties of the diagonal structures in Section 4.5 and the above assumptions about the transformation to physical scales, the diagonals correspond to an angle of about 65{\degs} relative to the line of sight. We therefore expect an overall structure to extend northeast from about zero velocity, and into the plane of the sky at an angle of 65{\degs} relative to the line of sight. The structure crosses the $xy$ plane roughly at the position of the ``blob'', and the position angle relative to the sky is PA$\simeq\!60\degs$. These angles should only be considered as guidelines to the orientation of the overall remnant structure, as the three main velocity structures in Fig.~\ref{fig:cubes} (middle and bottom panels) are extended in both the $x$ and the $y$ directions; this leads to some uncertainty in the exact orientation of the diagonal structure.

To better interpret the remnant structure, we used the visualisation tool \textsc{MayaVi}. We created interactive three-dimensional maps with this tool, using \textsc{Python} in an approach that follows \citet{Vogt10} for SNR~1E~0102.2-7219 in the Small Magellanic Cloud and \citet{Vogt11} for SNR~N132D in the LMC. The data cube was read in a four-dimensional array and was visualised point-by-point as partly transparent spheres; data of columns that contain spectra of broken fibres were removed in advance. The flux of each point is represented by both size and colour. A brighter colour and a larger size of a sphere correspond to a higher flux at a specific point. Foreground spheres are somewhat transparent to allow background spheres to shine through. We show [\ion{O}{iii}] and [\ion{S}{ii}] structures for three projections in Fig.~\ref{fig:mayavi}. To discriminate between approaching and receding gas, we colour-coded the gas differently, with a blue colour bar for approaching gas -- i.e.\ positive $z$ -- and a red colour bar for receding gas. Zero velocity corresponds to the local LMC rest velocity. We emphasise that the visualisation is mainly meant for highlighting structures, it must not be used in quantitative measurements.

We used \textsc{MayaVi} in three steps to produce the panels in Fig.~\ref{fig:mayavi}. In the upper panels of Fig.~\ref{fig:mayavi} we show scenes that are similar to those in the upper panels of Figs.~\ref{fig:full_images} and~\ref{fig:cubes}. Intensities for negative and positive $z$-values were produced separately. We also visualise the fainter flux around the central part separately, see Fig.~\ref{fig:mayavi2}; this component has similar colour coding. The scenes in the lower two panels of Fig.~\ref{fig:cubes} appear smaller in [\ion{S}{ii}] than in [\ion{O}{iii}]. This is merely an effect of the shorter velocity interval where we could measure line emission in [\ion{S}{ii}].

The right-hand side middle panel in Fig.~\ref{fig:mayavi} shows an overall diagonal structure that extends in the plane of the sky from the ``blob'' position at positive $x$-values at large inclination angle with the $z$-axis. This recovers the diagonal with 65{\degs} relative to the line of sight that we inferred from Fig.~\ref{fig:cubes} (cf. above). The structure appears to form two ellipse-like structures in [\ion{O}{iii}], one at about zero velocity, and the other towards the lower left corner in the middle left-hand side panel. The two ellipses are plausibly ring-like features at the ends of a bipolar structure. According to the velocities in Fig.~\ref{fig:cubes}, the upper right-hand side structure lies further in along the $y$-axis than the lower left-hand side structure. The [\ion{S}{ii}] structure is more complex; the $v\!\approx\!\pm400\kms$ structures, which we mention at the end of in Section 4.5, are clearly visible as the global diagonal structure and differ from the [\ion{O}{iii}] structures. In the bottom panels of Fig.~\ref{fig:mayavi} we view the possibly bipolar structure closer to its axis as viewed in the plane of the sky with the $x$-axis oriented inwards from East to West; different parts overlap more, and it is more difficult to disentangle individual features.

In Fig.~\ref{fig:mayavi2} we note abundant high-velocity gas in [\ion{O}{iii}] that fills the boxes in all panels; in particular on the near side of the box, as seen from the Earth (blue colour coding). This corresponds to the high-velocity shoulder that is seen in the line profiles in Fig.~\ref{fig:blobspec}.

\subsection{The structure of the pulsar wind nebula}
The large structure that we identified in the previous subsection and in Figs.~\ref{fig:cubes} and~\ref{fig:mayavi} has a position angle relative to the sky of PA$\simeq\!60\degs$. There is certainly some uncertainty to this position angle that might be high enough to accommodate the symmetry axis that is discussed in \paperone. In this previous study we find that many parameters change along this symmetry axis, and we argue for present and possible past shock activity along it; the symmetry-axis position angle is PA=45\degs--50\degs.

The ``blob'' region stands out in Figs.~\ref{fig:cubes} and~\ref{fig:mayavi}. An explanation to its existence could be a shock that propagates at $v\!\approx\!\pm400\kms$ along the line of sight. For normal cosmic abundances, the main problem with this interpretation is that cooling and recombination times are long for densities that are typical of this object. The post-shock temperature would be about $2\!\times\!10^6$K, and cooling times would be comparable to the age of the remnant for a post-shock electron density of $10^2\text{cm}^{-3}$. The situation is, however, different for {\snr} that has a high metal abundance. We have used our plasma code that is discussed in \citet{Sor04} and \citet{Mat08} to estimate cooling times for a range of abundances. In particular, with the relative abundances suggested by \citet{Williams08} for the oxygen-rich filaments -- i.e., O: Ne: Mg: Si: S: Ar: Ca: Fe = 1: 0.2: 0.1: 0.1: 0.1: 0.1: 0.1: 0.1 by number -- the cooling time scale is less than a few years for a post-shock electron density of $10\,\text{cm}^{-3}$ and temperatures corresponding to the shock velocities we tested, i.e., between 50--500\kms. We can even allow H and He to dominate, with H: He: O = 10: 2: 1 -- with all other metal abundances as in the oxygen-rich case -- and still have a cooling time scale of less than 10 years for a post-shock electron density of $10\,\text{cm}^{-3}$. It therefore seems plausible that a radiative shock that propagates at $v\!\approx\!\pm400\kms$ along the line of sight could exist at the ``blob'' position, but we cannot rule out slower shocks than this; if they are driven into already expanding structures, such as the ring-like features that are mainly seen in [\ion{O}{iii}]. Observations that sample [\ion{O}{ii}] emission at high spatial resolution would provide useful data to constrain the shock velocity. Unfortunately, VIMOS/IFU does not contain [\ion{O}{ii}]$\,\lambda\lambda3726,3729$; previous observations show that these lines are strong in the central region of {\snr} (\citealt{Kirshner89}; \paperser).

To explain shocks that propagate at velocities of up to 400{\kms} along the line of sight, we need to use similar arguments as \paperone; i.e., some event occurred at the ``blob'' position, which caused shocks to be driven into filaments there. The high $I_{\text{[\ion{S}{ii}]}}/I_{\text{[\ion{O}{iii}]}}$ ratio is not due to high local sulphur overabundance, if [\ion{S}{ii}] is concentrated to this location. The reason to the local release of energy at this location still needs to be explained, although energy input from pulsar activity seems plausible.

The faint [\ion{O}{iii}] emission in Fig.~\ref{fig:mayavi2} seems stronger on the near side of remnant than on the far side. Comparing the lower right-hand side panel of Fig.~\ref{fig:mayavi2} with the lower left-hand side panel of Fig.~\ref{fig:mayavi} it appears as if the faint [\ion{O}{iii}] emission is particularly strong at large $y$-values on the near side and small $y$-values on the far side. This could signal some asymmetry in the presumably photoionised ejecta in elongated structures along the jet axis, as in the Crab, or in ejecta beyond the PWN/ejecta interface.

\subsubsection{A discussion of the pulsar-activity axis}
Here we discuss a possible common origin to the structure we find here and the structures of \paperone. As we mentioned in the previous section, we assumed that the ring-like structures -- that are seen in Fig.~\ref{fig:mayavi} -- expand linearly. The position of the pulsar is $(x,y)\!\simeq\!(19,17)$, which corresponds to the position projected on the sky \citep{Seraf04}. We are tempted to suggest that the pulsar is positioned close to an axis that is put between the centres of the two ring-like features. This places the pulsar at a velocity along the line of sight of $v\!\simeq\!+400\kms$. A 3$\sigma$ upper limit on the proper motion on the sky is $290\kms$ \citep{DeLuca07}, which implies that the true pulsar motion is 400--$500\kms$, directed away from us at an angle that is within $\simeq\!36\degs$ of the line of sight. An argument against the pulsar sitting on the far (red) side of the nebula is the structure of faint [\ion{O}{iii}] emission (Fig.~\ref{fig:mayavi2}), which appears fairly centred on the pulsar at the LMC rest velocity.

In Fig.~\ref{fig:mayavi} [\ion{S}{ii}] appears to fill the structure more along the axis between the ring-like features than [\ion{O}{iii}]. If this axis is close to a pulsar-activity axis, such as a pulsar jet, or a part of a torus (see below), there is the possibility that some of the [\ion{S}{ii}]-emitting structures could be more affected by the pulsar than structures that primarily emit [\ion{O}{iii}]. The ``blob'' could be such a [\ion{S}{ii}]-emitting structure that was shocked by the pulsar activity. In such a scenario -- where past and present shocks are allowed along the pulsar-activity axis (\paperone) -- it is occasionally necessary to bring in gas towards this axis, or the pulsar input may vary along the activity axis. \citet{Buc04} and \citet{SG07} discuss processes that create and pull filaments into the PWN, but it is unclear how efficient such a filamentation is, and how late in the PWN evolution filaments could develop to remain unsuppressed. If it is the pulsar activity that varies, filaments may not have to move towards the axis. In this case activities in the PWN along the pulsar-activity axis vary instead.

\subsubsection{A comparison with the Crab using a torus model}
In a comparison with the Crab PWN \citet{Mori04} show that the outermost parts of the Crab PWN X-ray torus reach beyond 50{\arcsec} away from the pulsar. The outermost regions of the PWN torus of {\psr} (if similar to that of the Crab) would reside at $\gsim 2${\arcsec} from {\psr}, if we assume a distance of 2\,kpc to the Crab and 50\,kpc to \snr. \citet{Mori04} identify the entire torus in the Crab PWN to be external to the PWN termination shock, so the brightest parts of the X-ray torus should be in the post-shock flow. In particular, one can identify a bright feature at about 30{\arcsec} -- which corresponds to about $1\farcs2$ at the distance of \snr -- southwest of the Crab pulsar, which also shows the hardest X-ray spectrum in the PWN \citep[the photon index $\lsim\!1.8$, cf.\ fig~3 in][]{Mori04}. This is reminiscent of the situation in the PWN of {\psr}. In {\paperone} we find that the region close to the ``blob'', $1\farcs3$--$1\farcs4$ southwest of the pulsar, shows the hardest X-ray spectrum with a photon index of about 1.65. In the Crab, this region is on the far side of the torus, and is therefore not Doppler boosted. If we identify the ``blob'' region in the PWN of {\psr} in the same way -- that is, as an active region in the PWN torus -- we can speculate that the PWN torus in {\snr} could be Doppler boosted on the south-western side. The X-ray emission is namely brighter on that side compared to the north-eastern side. X-ray emission could also be enhanced due to interaction with supernova ejecta.

If we identify the symmetry axis in the PWN of {\snr} as a torus, we note that the difference between it and the Crab is that the optical line emission follows the X-ray emission closely in \snr. There is no such strong correlation in the Crab \citep[see, for example, fig.~11 in][]{Cad04}. This observation, and the in general higher temperatures that are derived from [\ion{O}{iii}] in the PWN of {\snr} compared to the Crab, favour a scenario where optical emission to a larger extent is driven by shocks rather than by photoionization in the PWN of {\snr}. The true ejecta distribution in the latter object may therefore not be correctly described by the structure as visualised in Figs.~\ref{fig:mayavi} and~\ref{fig:mayavi2}, since regions that are not excited by shocks will be fainter. The wide wavelength-range observational study of \citet{Charlebois2010} of the entire region of the Crab can provide a clue to the overall structure also for the PWN of \snr. Their three-dimensional movie shows bright loops that are visible preferentially in the torus plane, whereas structures in the jet directions are less structured and fainter. If the distance between the torus and the loops were shorter, a situation may arise such as that in the PWN of {\snr}. The discussion in \citet{Williams08} that oxygen-rich ejecta in {\snr} may have high-density structures that lag behind the general PWN/ejecta interface fits such a scenario.

\subsubsection{A comparison with Cas~A, SN~1987A, and 3D models}
If the ejecta structures we observe in [\ion{S}{ii}] and [\ion{O}{iii}] correspond to the real matter distribution -- and not just an effect of excitation by, for example, a PWN torus -- this could lead to clues about the supernova explosion. A parallel case with apparent asymmetric ejecta structures is SN~1987A, where observed inner ejecta lie in a plane that is close to the equatorial plane of the circumstellar ring \citep{Larsson2013}. Here structures of the metal-rich ejecta can be probed out to about 3500\kms, as a consequence of still active radioactive decay and ongoing interaction with the circumstellar ring. This higher velocity limit is twice as high as what we achieve with our VIMOS data for {\snr}. A direct comparison with SN~1987A is therefore not straightforward.

Another young well-observed core-collapse supernova remnant is Cas~A  \citep{Fesen2006,Delaney2010,Hwang2012}. A comparison with {\snr} is also in this case complicated, as the progenitor of Cas~A was a type IIb SN \citep{Krause2008} with an ejecta mass of perhaps as low as $3\Msun$ \citep{Hwang2012}. The ejecta of {\snr} are more massive, probably more similar to the ejecta mass of SN~1987A (\paperone). Due to the low ejecta mass, in combination with pre-supernova mass loss, the observed ejecta structures in Cas~A are affected by the reverse shock; the reverse shock has now reached freely expanding ejecta at about 5000{\kms} \citep{Hwang2012}. A massive asymmetric circumstellar shell is needed if a reverse shock were to be observationally important also in \snr. If we envision a circumstellar disk, as in SN~1987A, this may still leave an imprint if it was massive enough. A scenario with a circumstellar ring in {\snr} is discussed by \citet{Caraveo1998}, but it is not clear why it would be oxygen-rich, which is needed to explain the line emission in {\snr} (see, for example, fig.~6 in \pmorse).

The most advanced three-dimensional supernova simulation that is available for the full evolution from explosion until shock breakout is that of \citet{Hammer10}. It is based on a 15$\Msun$ blue supergiant progenitor model of \citet[see also \citealt{Kifonidis03}]{Woosley88}. Such a low-mass compact progenitor is most likely not a perfect model for the progenitor of \snr. In addition, the three-dimensional simulations do not include radioactive heating, which underestimates the volume of the radioactive elements. For {\snr} there is also the interaction between ejecta and the PWN, which alters the original structure. Nevertheless, we can probably draw some general conclusions on the three-dimensional structure of elemental abundances. The simulations clearly show how high-Z elements such as nickel ($^{56}$Ni) -- as well as other iron-group elements and silicon -- are mixed far out into the envelope, and are actually more mixed than oxygen, neon, and magnesium, on average. Looking in more detail, there are, however, large differences in the three-dimensional distribution of the elements. Most spectacular is the high-Z element distribution, with rather few long fingers protruding into the H/He envelope. This naturally explains the necessity of artificial mixing to large velocities of $^{56}$Ni in models of SN~1987A to explain its light curve  \citep[e.g.,][]{Blinn00}. Also, hydrogen is mixed in to low velocities ($\leq 3000 \kms$), which can explain why we see hydrogen in {\snr} (\paperser).

We expect to take the study presented here further in a future paper. With a full analysis of all spectra in Table~1, we expect to disentangle the three-dimensional structure of {\snr} at a higher signal-to-noise ratio, which will enable us to obtain a full three-dimensional map of the [\ion{S}{ii}] density, and not just projected densities. We will also include more elements, such as argon, to constrain the elemental distribution. This will allow us to make a more detailed comparison between the derived three-dimensional structure of the inner ejecta of {\snr} with models such as that of \citet{Hammer10}. Simulations using more relevant progenitors would naturally support this comparison.

\section{Conclusions}
We have presented and discussed new visual wavelength-range observations of the innermost region of the supernova remnant {\snr} in the LMC. We have used the integral-field spectrograph VIMOS/IFU that is mounted on the third unit telescope at the VLT of ESO in Chile. The IFU technique has up to now only been sparsely used with supernova remnants \citep[cf., for example,][]{Zharikov01}. Our observations have provided us with three-dimensional maps of [\ion{O}{iii}]$\,\lambda5007$ and [\ion{S}{ii}]$\,\lambda\lambda6717,6731$ at an $0\farcs33\!\times\!0\farcs33$ spatial sampling and a velocity resolution of about 31\kms. We have decomposed the sulphur lines and created a ``master'' [\ion{S}{ii}] line profile for each spectrum at high signal-to-noise ratio. We have also decomposed the oxygen lines to better probe high-velocity structures. In total we have about 1400 spectra with good [\ion{O}{iii}] and [\ion{S}{ii}] line profiles. We have not only made intensity maps for [\ion{O}{iii}] and [\ion{S}{ii}], but also for the intensity ratio $I_{\text{[\ion{S}{ii}]}}/I_{\text{[\ion{O}{iii}]}}$, and the average electron density.

We have, moreover, recovered previous results of \paperser, \pmorse, and {\paperone} in terms of the ``blob'', line profiles, and overall projected structures on the sky. Additionally, with our VIMOS/IFU data we have obtained more detailed information. Assuming linear expansion of the ejecta, we have created a three-dimensional model, which reveals a structure that stretches from the position of the ``blob'' (\paperone) and into the plane of the sky, at a position angle of PA$\simeq\!60\degs$. Assuming a remnant age of 1000 years and an object distance of 51\,kpc, the structure has an inclination angle of about $65\degs$ relative to the line of sight. The position angle is close to the symmetry axis of PA=45\degs--50\degs, which is sugested and discussed in {\paperone} to indicate present and past activities in the PWN. We speculate that the pulsar is positioned on this activity axis; in this case the pulsar has a velocity along the line of sight of a few hundred \!\kms.

The ``blob'' is most likely a region of shock activity, as it mainly emits in [\ion{S}{ii}]. A decisive observation to test this speculation is to make velocity maps also of an ion of another element, viz.\ [\ion{O}{ii}]$\,\lambda\lambda3726,3729$. The elongated structure of the central emission is, furthermore, embedded in a volume of diffuse [\ion{O}{iii}] emission, which gives rise to a wide shoulder of a spatially integrated line profile for the full field. The shoulder is somewhat brighter towards bluer wavelengths compared to redder wavelengths. So, whilst ejecta mainly emit on the far side of the remnant, diffuse emission is biased towards the near side. 

We have, finally, compared {\snr} to other supernova remnants and to a three-dimensional supernova model of \citet{Hammer10}. In particular, the striking resemblance in X-rays between the PWN of {\snr} and the Crab -- in combination with our findings in this paper -- suggests that the symmetry axis is part of a torus in the PWN, which is in agreement with \citet{GW00}, and not a pulsar jet as was later suggested. Contrary to for the Crab, supernova ejecta are more affected by the torus activity in \snr. The ``blob'' could be an unusually active part of such an interaction.

\section*{Acknowledgments}
We express our gratitude to Fr\'ed\'eric Vogt for his kind introduction to \textsc{MayaVi} as well as assistance with \textsc{Python} scripting. We thank Robert Fesen for valuable discussions on SNRs and Robert Cumming for data used in Fig.~\ref{fig:psr}. C.S.\ was supported by funds of DFG and Land Brandenburg (SAW funds from WGL), and also by funds of PTDESY-05A12BA1. P.L.\ acknowledges support from the Swedish Research Council, and G.O.\ support from the Swedish National Space Board. Y.S.  was partially supported by the RFBR (grants 11-02-00253 and 11-02-12082), RF Presidential Program (Grant NSh 4035.2012.2), and the Ministry of
Education and Science of the Russian Federation (Contract No. 11.G34.31.0001 and Agreement No.8409, 2012).

{}

\label{lastpage}

\begin{thebibliography}{}
\bibitem[\protect\citeauthoryear{Blinnikov \etal}{2000}]{Blinn00} 
Blinnikov S., Lundqvist P., Bartunov O., et\,al. 2000, ApJ, 532, 1132
 
\bibitem[\protect\citeauthoryear{Bucciantini \etal}{2004}]{Buc04} 
{Bucciantini} N., {Amato} E., {Bandiera} R., et\,al.\ 2004, \aap, 423, 253

\bibitem[\protect\citeauthoryear{Caraveo \etal}{1998}]{Caraveo1998}
{Caraveo} P., {Mignani} R., {Bignami} G.~F.\ 1998, \memsai, 69, 1061

\bibitem[\protect\citeauthoryear{\u{C}ade\u{z} \etal}{2004}]{Cad04} 
{\u{C}ade\u{z}} A., {Carrami\~nana} A., {Vidrih} S.\ 2004, ApJ, 609, 797

\bibitem[\protect\citeauthoryear{Charlebois \etal}{2010}]{Charlebois2010} 
{Charlebois} M., {Drissen}, L., {Bernier} A.-P.,  et\,al.\ 2010, AJ, 139, 2083

\bibitem[\protect\citeauthoryear{DeLaney \etal}{2010}]{Delaney2010}
{DeLaney} T., {Rudnick} L., {Stage} M. D., et\,al.\ 2010, ApJ, 725, 2038 

\bibitem[\protect\citeauthoryear{de Luca \etal}{2007}]{DeLuca07} 
{de Luca} A., {Mignani} R.~P., {Caraveo} P.~A., {Bignami} G.~F.\ 2007, ApJ, 667, 77

\bibitem[\protect\citeauthoryear{Fesen \etal}{2006}]{Fesen2006}
 {Fesen} R.~A., {Hammell} M.~C., {Morse} J., et\,al.\ 2006, 
 ApJ, 645, 283 

\bibitem[\protect\citeauthoryear{Gotthelf \& Wang}{2000}]{GW00} 
{Gotthelf} E.~V., {Wang} Q.~D.\ 2000, ApJ, 532, L117

\bibitem[\protect\citeauthoryear{Gr\"oningsson \etal}{2008}]{Gron08} 
{Gr\"oningsson} P., {Fransson} C., {Lundqvist} P., et\,al.\ 2008, A\&A, 479, 761

\bibitem[\protect\citeauthoryear{Hammer \etal}{2010}]{Hammer10} 
{Hammer} N.~J., {Janka} H.-Th., {M\"uller} E., 2010, ApJ, 714, 1371

\bibitem[\protect\citeauthoryear{Hester \etal}{2002}]{Hester02} 
{Hester} J.~J., {Mori} K., {Burrows} D., et\,al.\ 2002, ApJ, 577, 49

\bibitem[\protect\citeauthoryear{Hester}{2008}]{Hester08} 
{Hester} J.~J.\ 2008, ARA\&A, 46, 127 

\bibitem[\protect\citeauthoryear{Husemann \etal}{2012}]{Husemann12}
{Husemann} B., {Kamann} S., {Sandin} C., et\,al.\ 2012, \aap, 545, 137

\bibitem[\protect\citeauthoryear{Hwang \& Laming}{2012}]{Hwang2012}
{Hwang} U., {Laming}, J.~M. 2012, ApJ, 746, 130

\bibitem[\protect\citeauthoryear{Kaplan \etal}{2008}]{Kaplan08} 
{Kaplan} D.~L., {Chatterjee} S., {Gaensler} B.~M., {Anderson} J.\ 2008, ApJ, 677, 1201

\bibitem[\protect\citeauthoryear{Kifonidis \etal}{2003}]{Kifonidis03} 
{Kifonidis} K., {Plewa} T., {Janka} H.-Th., {M\"uller} E.\ 2003, \aap, 408, 621

\bibitem[\protect\citeauthoryear{Kirshner \etal}{1989}]{Kirshner89} 
{Kirshner} R.~P., {Morse} J.~A., {Winkler} P.~F., {Blair} W.~P.\ 1989, ApJ, 342, 260

\bibitem[\protect\citeauthoryear{Krause \etal}{1989}]{Krause2008} 
{Krause} O., {Birkmann} S.~M., {Usuda} T., et\,al.\ 2008, Science, 320, 1195

\bibitem[\protect\citeauthoryear{Lagerholm \etal}{2012}]{Lag12} 
{Lagerholm} C., {Kuntschner} H., {Cappellari} M., et\,al.\ 2012, \aap, 541, A82

\bibitem[\protect\citeauthoryear{Larsson \etal}{2013}]{Larsson2013} 
{Larsson} J., {Fransson}, C., {Kjaer} K.,  et\,al.\ 2013, submitted to ApJ (arXiv:1212.5051)

\bibitem[\protect\citeauthoryear{Le F\`evre \etal}{2003}]{LeFevre03}
{Le F{\`e}vre} O., {Saisse} M., {Mancini} D., et\,al.\ 2003, in {Iye} M., {Moorwood}, A.~F.~M., eds, Instrument Design and Performance for Optical/Infrared Ground-based Telescopes, Proc.\ SPIE, 4841, 1670

\bibitem[\protect\citeauthoryear{Lundqvist \etal}{2011}]{Lun11} 
{Lundqvist} N., {Lundqvist} P., {Bj\"ornsson} C.-I., et\,al.\ 2011, MNRAS, 413, 611 (\paperone)

\bibitem[\protect\citeauthoryear{Lundqvist \& Tziamtzis}{2012}]{LunTzi12} 
{Lundqvist} P., {Tziamtzis} A.\ 2012, MNRAS, 423, 1571

\bibitem[\protect\citeauthoryear{Maran \etal}{2000}]{Mar00} 
{Maran} S.~P., {Sonneborn} G., {Pun} C.~S.~J., et\,al.\ 2000, ApJ, 545, 390

\bibitem[\protect\citeauthoryear{Mattila \etal}{2008}]{Mat08}
{Mattila} S., {Meikle} W.~P.~S., {Lundqvist} P., et\,al.\ 2008, MNRAS, 389, 141

\bibitem[\protect\citeauthoryear{Melatos \etal}{2005}]{Melatos05} 
{Melatos} A., {Scheltus} D., {Whiting} M.~T., et\,al.\ 2005, ApJ, 633, 931 

\bibitem[\protect\citeauthoryear{Mori \etal}{2004}]{Mori04} 
{Mori} K., {Burrows} D.~N., {Hester} J.~J., et\,al. 2004, ApJ, 609, 186

\bibitem[\protect\citeauthoryear{Morse \etal}{2006}]{Morse06}
{Morse} J.~A., {Smith} N., {Blair} W.~P., et\,al.\ 2006, ApJ, 644, 188 (\pmorse)

\bibitem[\protect\citeauthoryear{Panagia}{2005}]{Panagia05} 
Panagia N.\ 2005, in Marcaide J.~M., Weiler K.~W., eds, IAU Coll.\ 192: Cosmic Explosions, On the 10th Anniversary of SN1993J, Springer Proc.\ Phys., 99, 585

\bibitem[\protect\citeauthoryear{Sandin \etal}{2010}]{San10} 
{Sandin} C., {Becker} T., {Roth} M.~M., et\,al.\ 2010, \aap, 515, 35

\bibitem[\protect\citeauthoryear{Sandin \etal}{2011}]{San11} 
{Sandin} C., {Weilbacher} P., {Streicher} O., et\,al.\ 2011, The Messenger, 144, 13

\bibitem[\protect\citeauthoryear{Sandin \etal}{2012}]{San12} 
{Sandin} C., {Weilbacher} P., {Tabataba-Vakili} F., {Streicher} O.\ 2012, in Radziwill N.~M., Chiozzi G., eds, Software and Cyberinfrastructure for Astronomy II, Proc. SPIE, 8451, 84510F

\bibitem[\protect\citeauthoryear{Serafimovich \etal}{2004}]{Seraf04} 
{Serafimovich} N.~I., {Shibanov} Y.~A., {Lundqvist} P., {Sollerman} J.\ 2004, \aap, 425, 1041

\bibitem[\protect\citeauthoryear{Serafimovich \etal}{2005}]{Seraf05} 
{Serafimovich} N.~I., {Lundqvist} P., {Shibanov} Y.~A., {Sollerman} J.\ 2005, AdSpR, 35, 1106 (\paperser)

\bibitem[\protect\citeauthoryear{Sharp \& Birchall}{2010}]{Sharp10}
{Sharp} R., {Birchall} M.~N.\ 2010, PASA, 27, 91

\bibitem[\protect\citeauthoryear{Sorokina \etal}{2004}]{Sor04}
{Sorokina} E.~I., {Blinnikov} S.~I., {Kosenko} D.~I., {Lundqvist} P. 2004, Astron.\ Lett., 30, 737

\bibitem[\protect\citeauthoryear{Stone \& Gardiner}{2007}]{SG07} 
{Stone} J.~M., {Gardiner} T.\ 2007, ApJ, 671, 1726

\bibitem[\protect\citeauthoryear{Trimble}{1968}]{Tri68} 
{Trimble} V.\ 1968, AJ, 73, 535

\bibitem[\protect\citeauthoryear{Vogt \& Dopita}{2010}]{Vogt10} 
 {Vogt} F., {Dopita} M.~A.\ 2010, ApJ, 721, 597
 
\bibitem[\protect\citeauthoryear{Vogt \& Dopita}{2011}]{Vogt11} 
 {Vogt} F., {Dopita} M.~A.\ 2011, Ap\&SS, 331, 521
 
\bibitem[\protect\citeauthoryear{Weisskopf \etal}{2000}]{Weisskopf00} 
{Weisskopf} M.~C., {Hester} J.~J., {Tennant} A.~F., et\,al.\ 2000, ApJ, 536L, 81
 
\bibitem[\protect\citeauthoryear{Williams \etal}{2008}]{Williams08} 
{Williams} B.~J., {Borkowski} K.~J., {Reynolds} S.~P., et\,al.\ 2008, ApJ, 687, 1054

\bibitem[\protect\citeauthoryear{Williams}{2010}]{Williams10} 
{Williams} B.~J.\ 2010, PhD Thesis (North Carolina State Univ., Raleigh)
[arXiv:1005.1296]

\bibitem[\protect\citeauthoryear{Woosley \etal}{1988}]{Woosley88} 
{Woosley} S.~E., {Pinto} P.~A., {Ensman} L.\ 1988, ApJ, 324, 466

\bibitem[\protect\citeauthoryear{Zharikov \etal}{2001}]{Zharikov01} 
{Zharikov} S., {Shibanov} Yu., {Koptsevich} A., et\,al.\ 2001, in {Lee} H., {Torres-Peimbert} S., eds., The Seventh Texas-M\'exico Conf.\ on Astroph.: Flows, Blows and Glows, RevMexAA Conf.\ Ser., 10, 163

\end{thebibliography}
\end{document}